\documentclass[preprint,10pt]{aastex}
\usepackage{epsf}
\newcommand{\al}{\alpha}
\newcommand{\hMpc}{h^{-1}{\rm\;Mpc}}

\newcommand{\trihGpc}{h^{-3}{\rm\;Gpc^3}}
\newcommand{\trihMpc}{h^{-3}{\rm\;Mpc^{3}}}

\newcommand{\ihMpc}{h{\rm\;Mpc^{-1}}}
\newcommand{\Mpc}{{\rm\;Mpc}}

\newcommand{\kpe}{k_\perp}
\newcommand{\kpa}{k_\parallel}
\newcommand{\spe}{s_\perp}
\newcommand{\spa}{s_\parallel}
\newcommand{\Om}{\Omega_m}
\newcommand{\Obb}{\Omega_b}
\newcommand{\Occ}{\Omega_c}

\newcommand{\OL}{\Omega_\Lambda} 
\newcommand{\Ok}{\Omega_K}
\newcommand{\DA}{D\!_A(z)}
\newcommand{\hz}{H(z)}
\newcommand{\DAA}{D_A} 
\newcommand{\hzz}{H}

\newcommand{\Vsur}{V_{\rm survey}}
\newcommand{\Veff}{V_{\rm eff}}

\newcommand{\Neff}{N_{\rm eff}}
\newcommand{\Oh}{\Omega_m h^2}
\newcommand{\Ob}{\Omega_b h^2}
\newcommand{\kmax}{k_{\rm max}}
\newcommand{\kfit}{k_{\rm fit}}
\newcommand{\PBlin}{P_{b,{\rm lin}}}
\newcommand{\PBnl}{P_{ b,{\rm nl}}}
\newcommand{\PC}{P_{ c}}
\newcommand{\PB}{P_{ b}}
\newcommand{\Plin}{P_{\rm lin}}
\newcommand{\Sig}{\Sigma}
\newcommand{\sig}{\sigma}
\newcommand{\Sigs}{\Sigma_s}
\newcommand{\Sign}{\Sigma_{\rm nl}}
\newcommand{\Sigpa}{\Sigma_\parallel}
\newcommand{\Sigpe}{\Sigma_\perp}
\newcommand{\Sigz}{\Sigma_z}

\newcommand{\kref}{k_{\rm ref}}

\newcommand{\nPt}{nP_{0.2}}
\newcommand{\Pt}{P_{0.2}}
\newcommand{\sn}{n^{-1}}
\newcommand{\roff}{r}

\newcommand{\ks}{k_{\rm silk}}
\newcommand{\so}{s_o}
\begin{document}
%\title{Improved forecasts for the baryon acoustic oscillations and cosmological distance scale}
%
\title{IMPROVED FORECASTS FOR THE BARYON ACOUSTIC OSCILLATIONS AND COSMOLOGICAL DISTANCE SCALE}
\author{Hee-Jong Seo \& Daniel J. Eisenstein}
\affil{Steward Observatory, University of Arizona, 933 North Cherry Avenue, Tucson, AZ 85721}
\email{hseo@as.arizona.edu,deisenstein@as.arizona.edu}
\affil{Submitted to \textit{The Astrophysical Journal} 12-20-2006}
%{Submitted to \textit{The Astrophysical Journal} 12-20-2006} 
\begin{abstract}

We present the cosmological distance errors achievable using the
baryon acoustic oscillations as a standard ruler. We begin from
a Fisher matrix formalism that is upgraded from \citet{SE03}. We
isolate the information from the baryonic peaks by excluding distance
information from other less robust sources. Meanwhile we accommodate the
Lagrangian displacement distribution into the Fisher matrix calculation
to reflect the gradual loss of information in scale and in time due to
nonlinear growth, nonlinear bias, and nonlinear redshift distortions.
We then show that we can contract the multi-dimensional Fisher matrix
calculations into a 2-dimensional or even 1-dimensional formalism with
physically motivated approximations. We present the resulting fitting
formula for the cosmological distance errors from galaxy redshift surveys
as a function of survey parameters and nonlinearity, which saves us going
through the 12-dimensional Fisher matrix calculations. Finally, we show excellent agreement between the distance error estimates from the revised Fisher matrix and the precision on the distance scale recovered from N-body simulations.
\end{abstract}

\section{Introduction}
The famous Hubble expansion drives more distant objects to recede faster from us. Recent observations of supernovae argue that this expansion is in fact accelerating, implying an existence of dark energy with negative pressure \citep[][]{Riess98,Perlm99,Rie01,Knop03,Ton03,Riess04}. This dark energy, which contributes two third of energy density in the present Universe, is mysterious in its physical origin. Precise measurements of its time evolution will be crucial to uncover the identity of this energy component. One of most promising probes to measure the dark energy is a standard ruler called baryon acoustic oscillations in large-scale clustering \citep[][]{Eht98,Eht99}

Baryon acoustic oscillations (hereafter BAO) arise from sound waves that propagated in the hot plasma of tightly coupled photons and baryons in the early Universe. As the Universe expands and cools, photons finally decouple from baryons 400,000 years after the Big Bang, with the sound waves revealed as the acoustic oscillations in the anisotropies of the Cosmic Microwave Background (hereafter CMB) \citep{Mil99,deB00,Han00,Lee01,HalDasi,Netter02,BenoitArcheops,BennettWmap,Pearson03,Hinshaw06}. The equivalent but attenuated feature exists in the clustering of matter, as baryons fall into dark matter potential well after the recombination \citep{Peebles70,SZ70,Bon84,Holtzman89,HS96,EH98}. These were recently detected in galaxy redshift surveys \citep{Eisen05,Cole05,Hutsi06,Percival06,Tegmark06}. Good CMB anisotropy data provides the absolute physical scale for these baryonic peaks; comparing this to the observed location of the baryonic peaks in the two-point correlation function or power spectrum provides measurements of cosmological distance scale. Clustering in the transverse direction probes the angular diameter distance, and clustering in the line-of-sight direction probes the Hubble parameter. The cosmological distance scale as a function of redshift is then the record of the expansion history of the Universe, which in turn measures the evolution of dark energy density.   

Many studies have been devoted to estimations of precision on cosmological distance scale achievable using the BAO from future galaxy redshift surveys \citep{Eisen03,Blake03,Linder03,Hu03,SE03,Cooray04,Mat04,Amen05,Blake05,Glazebrook05,Dolney06,Zhan06,Blake06}. A common scheme is to assume the Gaussian statistical errors on the power spectrum, which are constructed straightforward from the finite volume of the survey and the incomplete sampling of the underlying density field with galaxies \citep{Teg97}, and then to propagate these errors to constrain the errors on cosmological parameters including cosmological distance scale whether by using Monte-Carlo simulations or using analytic method such as a Fisher matrix formalism. In \citet{SE03}, we used a Fisher matrix formalism to show that the BAO from future galaxy redshift surveys can constrain the cosmological distance scale to percent level precision, thereby providing robust measurements of present-day dark energy density and its time-dependence that are comparable to future Type Ia supernova surveys.

The most important element in these methods, other than the survey specifications, is how to include the effects of nonlinear structure formation, which depends on redshift as well as bias. The primary nonlinear effect on the standard ruler test is that the BAO signature is reduced with time by the nonlinear growth of density fields into present-day structures \citep{Meiksin99,Springel05,Angulo05,SE05,White05,Crocce06,Jeong06,Huff06}. This nonlinear mode-coupling also alters the overall shape of the power spectrum by increasing the small-scale power relative to the linear growth of power spectrum \citep[][and references therein]{Jain94}. At the same time, this mode-coupling increases statistical errors above our Gaussian approximations \citep{MW99,Sco99}. The nonlinear effects first appear on small scales and then proceed to larger scales, tracing the hierarchical formation of structures. Nonlinear redshift distortions and nonlinear galaxy bias can further degrade the BAO \citep[e.g.,][]{Springel05,Angulo05,SE05,White05,Huff06} and modify the statistical variance.

In \citet{SE03}, we used a sharp wavenumber cut to exclude the nonlinear wavenumbers from the Fisher matrix formalism. That is, we treated all wavenumbers up to a certain threshold as linear and excluded all larger wavenumbers from our calculations, while the threshold varied depending on redshift. Our choice of linear wavenumbers was based on the magnitude of underlying mass density fluctuations rather than direct estimations of degradation of baryonic signature with redshift, as we did not have any means to estimate the latter without N-body simulations. In addition, the effects of bias or redshift distortions on the BAO were too ambiguous to parameterize and include in the Fisher matrix. However, the nonlinear effect does not turn on sharply with wavenumber but instead appears gradually. In \citet{SE05}, we used N-body simulations to study the effect of nonlinear growth, bias, and redshift distortions on the appearance of the BAO. While we have shown in that paper that our choice of linear scales in \citet{SE03} reasonably represents the amount of standard ruler information at different redshift, we still lacked a quantitative description of the gradual erasure on the BAO as a function of redshift, bias, and redshift distortions. 

Recently, a quantitative model has been provided by \citet{ELag06}. In a two-point correlation function, the BAO signature is an excess of pairs separated by its characteristic scale of $\sim 100\hMpc$. \citet{ELag06} models the nonlinear process of erasing the BAO signature as differential motions or Lagrangian displacements of these pairs of tracers. Using the N-body simulations from \citet{SE05}, we estimated the amount of differential motions caused by nonlinear growth, halo bias, and redshift distortions at different redshifts. We then showed that the nonlinear erasure of the BAO signature is successfully modeled by approximating these displacement field as Gaussian and convolving this Lagrangian displacement distribution with the BAO peak in the two-point correlation function, or multiplying the linear power spectrum by the Fourier transform of the Gaussian function. As noted in \citet{ELag06}, estimating the Lagrangian displacement fields with a reasonable precision does not require a simulation as large as estimating the degradation in the BAO signature does: Lagrangian displacements are calculated from all pairs separated by $\sim 100\hMpc$, while the BAO depend on the small number of excess pairs at that separation. 

In this paper, we use the Lagrangian displacement distribution calculated from \citet{ELag06} to modify the Fisher matrix formalism presented in \citet{SE03} so as to correctly reflect the gradual loss of standard ruler information from the BAO in scale and in time due to not only nonlinear growth but also bias and redshift distortions. We isolate the cosmological distance information from the BAO more strictly than in \citet{SE03}, i.e., without information from \citet{AP79} test or from the broad-band shape of power spectrum, and derive distance error estimates using the Fisher matrix calculations. We show that the full, 12-dimensional (for one redshift bin) Fisher matrix in \citet{SE03} can be contracted to a 2-dimensional or even 1-dimensional covariance matrix of $\DAA$ and $\hzz$ with physically motivated approximations. We present the resulting fitting formula, for errors from the full dimensional Fisher matrix, based on the reduced 2-dimensional covariance matrix, which is a function of survey volume, galaxy number density, and parameterized nonlinearity. Note that the effect of bias and redshift distortions on the BAO is now straightforward to parameterize as an increase in Lagrangian displacement field and easy to mingle into the Fisher matrix calculation. We show that the application of the formula extends to the photometric redshift surveys. This fitting formula will save considerable amount of computational efforts in forecasting distance errors for future galaxy surveys. We compare these results with distance constraints from a $\chi^2$ analysis of the full N-body results from \citet{SE05}. Our fitting formula differs from \citet{Blake06} in many details, such as the different treatments of nonlinear degradation of standard ruler test and ours having a structure based on physically motivated model.

In \S~\ref{sec:full}, we present the full Fisher matrix calculation that is upgraded by using the Lagrangian displacement field. In \S~\ref{sec:1DA}, we show a  1-dimensional model as an approximation of the full Fisher matrix in \S~\ref{sec:full}. In \S~\ref{sec:2DA}, the 1-dimensional model is extended to a 2-dimensional model. We present the resulting fitting formula. We also discuss the extension of this formula for photometric redshift surveys. In \S~\ref{sec:result}, we compare distance errors from the fitting formula with the estimates from the full Fisher matrix. In \S~\ref{sec:Nbody}, we compare the distance errors from the full Fisher matrix calculations (or from the fitting formula) with the distance errors from a $\chi^2$ analysis of the N-body data.

\section{Full Fisher matrix calculation}\label{sec:full}
In this section, we construct the Fisher matrix that assesses the amount of distance information available exclusively from using the BAO as a standard ruler. This enables us to use the Lagrangian displacement distribution \citep{ELag06} to account for the nonlinear effects on baryonic peaks and therefore nonlinear effects on the distance precision.

Assuming the likelihood function of the band powers of galaxy power spectrum to be Gaussian, the Fisher matrix is approximately \citep{Teg97,SE03}:
\begin{eqnarray}  
F_{ij}&=&\int_{\vec{k}_{\rm min}} ^ {\vec{k}_{\rm max}} \frac{\partial \ln P(\vec{k})}{\partial p_i} \frac{\partial \ln P(\vec{k})}{\partial p_j} \Veff(\vec{k})\frac{d\vec{k}}{2(2\pi)^3}
\label{eq:Fij} \\
&=&\int_{-1}^{1} \int_{k_{\rm min}}^{\kmax}\frac{\partial \ln P(k,\mu)}{\partial p_i} \frac{\partial \ln P(k,\mu)}{\partial p_j} \Veff(k,\mu) \frac{2\pi k^2 dk d\mu}{2(2 \pi)^3}  \nonumber 
\end{eqnarray}
where $P(\vec{k})$ is the observed power spectrum at $\vec{k}$, $\mu$ is the cosine of the angle of $\vec{k}$ with respect to the line-of-sight direction, $p_i$ are the cosmological parameters to be constrained, and $\Veff$ is the effective volume of the survey, given as 
\begin{equation}
\Veff(k,\mu) = 
\int \left [ \frac{{n}(\vec{r})P(k,\mu)}{{n}(\vec{r})P(k,\mu)+1} \right ]^2 d\vec{r} 
=\left [ \frac{{n}P(k,\mu)}{{n}P(k,\mu)+1} \right ]^2 \Vsur
=\left [ \frac{{n}P(k)(1+\beta \mu^2)^2}{{n}P(k)(1+\beta \mu^2)^2+1} \right ]^2 \Vsur.
\label{eq:Veff} 
\end{equation}
Here $n(\vec{r})$ is the comoving number density of galaxies at the location $\vec{r}$, $\beta$ is the linear redshift distortion parameter, and $\Vsur$ is the survey volume.
This formulation assumes that the density field is Gaussian and that boundary effects are not important. The second equality in equation (\ref{eq:Veff}) holds only if the comoving number density $n$ is constant in position, and the third equality assumes linear redshift distortions \citep{Kaiser87}.

From \citet{ELag06}, the Lagrangian displacement fields due to nonlinear
growth and nonlinear redshift distortions at separation of $100\hMpc$ can be approximated
as an elliptical Gaussian function. Since most of the effect
comes from the bulk flow, galaxy bias does not have much effect on Lagrangian displacement. The rms radial displacements across
($\Sigpe$) and along the line of sight ($\Sigpa$) at this separation
follow $\Sigpe=\Sig_0 G$ and $\Sigpa=\Sig_0 G(1+f)$, where $\Sig_0 =
12.4\hMpc$ for a cosmology with $\sigma_8=0.9$ at the present-day, 
$f=d(\ln G)/d(\ln a)\sim \Om^{0.6}$, and $G$ is the growth
function while $G$ is normalized to be $G=0.758$ at $z=0$ such that
$G(z)=(1+z)^{-1}$ at high $z$.  $\Sig_0$ should be scaled proportionally to 
$\sigma_8$.
Then the surviving baryonic features in the nonlinear 
power spectrum can be expressed as:
\begin{eqnarray} \label{eq:surPbao}
\PBnl(k,\mu)&=&\PBlin(k,\mu)\exp{\left (-\frac{\kpe^2\Sigpe^2}{2}-\frac{\kpa^2\Sigpa^2}{2} \right )} \\
&=&\PBlin(k,\mu)\exp{\left [ -k^2 \left (\frac{(1-\mu^2)\Sigpe^2}{2}+\frac{\mu^2\Sigpa^2}{2} \right )\right ]}
\end{eqnarray}
\noindent where $\PBlin$ is the portion of the linear power spectrum $\Plin$ with the acoustic signature, including the effects of Silk damping \citep{Silk68}.

From equations (\ref{eq:Fij}) and (\ref{eq:Veff}), the distance precision from a galaxy redshift survey depends on $\Vsur$, $nP$, and the redshift of the survey. The redshift dependence enters because of the loss of information due to nonlinear effect, such as the erasure of baryonic features as well as the nonlinear growth of power at large wavenumber. In \citet{SE03}, the scales degraded by nonlinear effects were removed from the Fisher matrix calculation by cutting off the wavenumber integral at a $\kmax$, where $\kmax$ depends on redshift by requiring $\sigma(r)\sim 0.5$ at a corresponding $r=\pi/2\kmax$. In reality, the nonlinear effect is progressive with $k$: there is acoustic scale information even beyond $\kmax$ while some information is lost within $k<\kmax$ \citep{SE05}. Now that we have a good model to describe the degradation in baryonic features (eq.\ [\ref{eq:surPbao}]), we can remove the redshift dependent $\kmax$. We set $\kmax =0.5\ihMpc$ for all redshifts and change $\Sigpa$ and $\Sigpe$ to account for the erasure of baryonic features.

As we seek to isolate the information from the acoustic scale, we attach the exponential suppression in equation (\ref{eq:surPbao}) to the full power spectrum when computing $\partial P(k,\mu)/\partial p_i$. This means that we are removing all information, baryonic and non-baryonic, from smaller scales, where the broad-band shape of the power spectrum might give distance information.  Moreover, when we compute the derivatives of $P$, we take the exponential factor in equation (\ref{eq:surPbao}) outside of the derivatives. This is equivalent to marginalizing over a large uncertainty in $\Sigpe$ and $\Sigpa$ when computing distance errors; we do not want to include any distance information from the anisotropy and scale of the nonlinear damping of the acoustic peaks.

With these prescriptions, the Fisher matrix becomes 
\begin{eqnarray}  \label{eq:fullFij}
F_{ij}
&=&\int_{-1}^{1} \int_{0}^{\infty}\frac{\partial \ln \Plin(k,\mu)}{\partial p_i} \frac{\partial \ln \Plin(k,\mu)}{\partial p_j} \Veff(k,\mu) \exp{\left [ -k^2\Sigpe^2-k^2\mu^2(\Sigpa^2-\Sigpe^2) \right ]}\frac{2\pi k^2 dk d\mu}{2(2 \pi)^3}\nonumber \\
\end{eqnarray}
where $\Plin(k,\mu)$ is the linear power spectrum.

To this point, we have not excluded the distance information on linear scales from non-baryonic features. There are two sources of such information: first, from the \citet{AP79} test (hereafter AP test) using angular anisotropy of power spectrum, and second, from the power spectrum not following a simple power law. 

First, at a given $k$, both the redshift distortions from large scale infall and the cosmological distortions give rise to angular anisotropies in power. When the linear redshift distortion parameter $\beta$ becomes very small, both distortions converge to an identical angular signature, a quadratic in $\mu$, causing the effect of redshift distortions to be degenerate from the effect of $\DAA\hzz$. In this case, the distance information from the AP test will be suppressed. However, when $\beta$ is non-negligible, as we assume a specific angular dependence of redshift distortions (e.g., linear redshift distortions), the two effects can be in principle distinguished as both distortions have different higher multiples, and we can therefore isolate the cosmological distance information on $\DAA \hzz$ (i.e., the AP test) from the effect of $\beta$. 

Second, if the galaxy power spectrum follows a simple power law without any preferred scale, the cosmological distortions remain degenerate from the redshift distortions for a negligible $\beta$. However, the galaxy power spectrum contains another preferred scale, the horizon scale at the epoch of matter-radiation equality. This effect produces much broader feature in power spectrum, but in principle can provide an extra standard ruler \citep{Cooray01}. 

However, we consider both of these sources of information less robust than the BAO. For the AP test, we do not believe we understand the quasilinear and nonlinear behavior of redshift distortions very accurately. For the broadband shape of the power spectrum, the resulting distance information will be susceptible to the systematic effect such as tilt, nonlinear bias, or nonlinear growth in power.

Therefore we want to remove distance information from both non-baryonic features.  For the distance information from the broad-band shape, we remove this by computing the Fisher matrix $F_{\Obb= 0.005}$ with $\Obb= 0.005$\footnote{We cannot use  $\Obb=0$ to produce a reasonable CMB information.} and subtracting it from the Fisher matrix $F_{ \Obb}$ with the fiducial $\Obb$. In addition, in order to remove any distance information from a presumed form of redshift distortions, we assume $\beta \sim 0$ in the derivatives in equation (\ref{eq:fullFij}) in computing both $F_{\Obb=0.005}$ and the fiducial Fisher matrix. That is, the angular AP test is removed. Meanwhile, we hold the amplitude of power in $\Veff$ and therefore the statistical error per each $k$ bin constant by keeping $\beta$ in the redshift distortions $R$ in $\Veff$ (eq.\ [\ref{eq:Veff}]) unchanged. Note that our inclusion of an arbitrary growth function already implies that any constraints on $\beta$ are not yielding distance scale information from the amplitude of the power spectrum; setting $\beta\sim 0$ in derivatives will not have any further effects. The resulting Fisher matrix after subtraction by $F_{\Obb=0.005}$ (while $\sig_{8,m}$ is the same) will only contain the distance information from the BAO. We ignore the minor effect that the broad-band shape of the fiducial power spectrum is slightly different from that of the $\Obb=0.005$ case.

The resulting Fisher matrix is combined with the CMB information ({\it Planck} satellite including polarization) and then inverted to give a covariance matrix. We calculate the distance errors marginalized over a total of 12 parameters including the angular diameter distance ($\ln D_A$) and the Hubble parameter ($\ln H$) at the redshift of the galaxy survey and 10 others: $\Oh$, $\Ob$, $\tau$, tilt ($n_s$), $T/S$, the normalization, $\ln {\DAA} ({\rm CMB})$, $\ln \beta$, an unknown growth rate ($G(z)$), and an unknown shot noise. 

In detail, we use the 1st year and the 3rd year Wilkinson Microwave Anisotropy Probe (hereafter, WMAP1 and WMAP3, respectively) results as our fiducial cosmologies \citep{Spergel03,Spergel06}. Our fiducial model for WMAP1 is then $\Om = 0.27$, $h=0.72$, $\OL=0.73$, $\Ok=0$, $\Ob = 0.0238$, $\tau = 0.17$, $n_s = 0.99$, and $T/S = 0$. For WMAP3, we use $\Om = 0.24$, $h=0.73$, $\OL=0.76$, $\Ok=0$, $\Ob = 0.0223$, $\tau = 0.09$, $n_s = 0.95$, and $T/S = 0$.

By adopting $\Sign$ (i.e., $\Sigpe$ and $\Sigpa$), we no longer need to specify the redshift of survey, and the distance error will only depend on $\Vsur$, $nP$, and $\Sign$. We assume a fiducial galaxy survey of $\Vsur=1\trihGpc$. While there is a simple scaling relation for varying $\Vsur$, the effect of $nP$ depends on the value of $\Sign$. We try various $nP$ and a large range of $\Sign$. We characterize $nP$ as a value at $k=0.2\ihMpc$, i.e., $\nPt$, where $P$ is real-space power. In redshift space, the linear theory redshift distortions enhance the power along the line-of-sight direction by $R=(1+\beta \mu^2)^2$ against shot noise.

\section{1-D approximation of Fisher matrix : a centroid problem}\label{sec:1DA}
Motivated by the single peak in the correlation function, we next consider whether acoustic distance scale precision can be modeled simply as the problem of centroiding a peak in the presence of the noise generated from shot noise and the CDM power spectrum. We study this in spherical geometry in this section and generalize to the anisotropic case in the next. Nonlinear effects will broaden the peak and therefore increase the uncertainty in measuring the location of the peak. We approximate the full Fisher matrix based on how well we can centroid the location of the peak,  that is, the sound horizon $\so$ at the drag epoch when observed in the reference cosmology:
\begin{eqnarray} \label{eq:Fii}
F_{\ln \so}&=&\Vsur\int_{k_{\rm min}}^{k_{\rm max}} \frac{1}{(P(k)+n^{-1})^2}\left [ \frac{\partial \PB(k)}{\partial \ln \so} \right ] ^2 \frac{4\pi k^2 dk}{2(2 \pi)^3}.   
\end{eqnarray}
We first find an appropriate form of $\partial \PB(k)/\partial \ln \so$. 
If the peak in the correlation function is a delta function at $r=\so$, the Fourier transformation of the peak takes the following form in the power spectrum:
\begin{equation}
\PB(k) \propto \frac{\sin{k\so}}{k\so}.
\end{equation}
A broadened peak will have the same power spectrum multiplied with an additional damping envelope set by the Fourier transformation of the broadened peak shape. For example, if the peak in the correlation function is broadened with a Gaussian function due to the Silk damping effect ($\Sigs$) and Lagrangian displacement distribution ($\Sign$), then the convolution in configuration space with the Gaussian function is Fourier-transformed to a multiplicative exponential factor in Fourier space. 
\begin{equation}\label{eq:gausilk}
\PB(k) \propto \frac{\sin{k\so}}{k\so}\exp{(-k^2 \Sig^2/2)}
\end{equation}
\noindent where $\Sig^2=\Sigs'^2+\Sign^2$. 

In reality, the decoupling between photons and baryons is not instantaneous at the recombination, and therefore the resulting Silk-damping effect $\mathcal{D}(k)$ integrated over time deviates from a Gaussian and can be better approximated by $\exp{(-(k/\ks)^{1.4})}$ with the Silk-damping scale $\ks(\equiv 1/\Sigs)$ \citep{HS96,EH98}. We therefore have 
\begin{equation}
\PB(k) \propto \frac{\sin{k\so}}{k\so}\exp{(-(k/\ks)^{1.4})}\exp{(-k^2 \Sign^2/2)}= \frac{\sin{k\so}}{k\so}\exp{(-(k\Sigs)^{1.4})}\exp{(-k^2 \Sign^2/2)}. \label{eq:ksilk}
\end{equation}

Starting from physical models of the transfer function gives the same results.
Let $\PB=P-\PC$. 
\begin{eqnarray}\label{eq:fullP}
P(k) &=& A k^nT^2(k)= A k^n\left [ \frac{\Obb}{\Om} T_b(k)+\frac{\Occ}{\Om} T_c(k) \right ]^2\\
&=&A k^n \left [ \left( \frac{\Occ}{\Om} T_c(k) \right )^2+ 2\frac{\Occ}{\Om}\frac{\Obb}{\Om} T_b(k)T_c(k)+\left( \frac{\Obb}{\Om} T_b(k) \right )^2 \right ]
\end{eqnarray}
As $kT_c(k)$ is a slow function of $k$ relative to $T_b(k)$, the leading order term for baryonic features can be approximated to
\begin{equation}
\PB(k) \propto 2\frac{\Occ}{\Om} \frac{\Obb}{\Om} T_b(k).
\end{equation}
From \citet{EH98} (and references therein), $T_b(k) \propto \mathcal{D}(k)\sin{k\so}/{(k\so)}$ where the Silk damping effect $\mathcal{D}(k)\sim \exp{(-(k/\ks)^{1.4}})\equiv \exp{(-(k\Sigs)^{1.4}})$. Including the nonlinear damping returns equation (\ref{eq:ksilk}).
\begin{equation}
\PB(k) \propto \frac{\sin{k\so}}{k\so}\exp{(-(k\Sigs)^{1.4})}\exp{(-k^2 \Sign^2/2)}.
\end{equation}
As the amplitude of $\PB(k)$ grows with redshift by $G^2$, we describe the normalization of $\PB$ as:
\begin{equation}\label{eq:Pbao} 
\PB(k) =\sqrt{8\pi^2} A_0 P_{0.2} \frac{\sin{k\so}}{k\so}\exp{(-(k\Sigs)^{1.4})}\exp{(-k^2 \Sign^2/2)} 
\end{equation}
\noindent where $P_{0.2}$ is galaxy power at $k=0.2\ihMpc$ at the given redshift.
We now differentiate $\PB(k)$ to calculate $\partial \PB(k)/\partial \ln \so$. 
\begin{equation}\label{eq:dPbao} 
\frac{\partial  \PB(k)}{\partial \ln \so}
= \sqrt{8 \pi^2} A_0 P_{0.2}\left[ \cos{k\so}-\frac{\sin{k\so}}{k\so} \right ] \exp{(-(k\Sigs)^{1.4})}\exp{(-k^2 \Sign^2/2)}.\end{equation}
Then the Fisher matrix becomes:
\begin{eqnarray}
F_{\ln \so}&=&\int_{0}^{\infty} \frac{8\pi^2 \Vsur  }{(P(k)+\sn)^2}
\left[ A_0 \Pt \left (\cos{k\so}-\frac{\sin{k\so}}{k\so} \right ) \right ]^2\exp{(-2 (k\Sigs)^{1.4})}\exp{(-k^2 \Sign^2)} \frac{4\pi k^2 dk}{2(2 \pi)^3}. \nonumber\\
\end{eqnarray}
The comoving sound horizon
is $\sim$ $100\hMpc$ and most of baryonic information comes from $k
\gtrsim 0.05\ihMpc$, which keeps $k\so$ large over the wavenumber of our
interest. Therefore, the sinusoidal terms oscillate rapidly relative to
the wavenumber dependence of the exponential factor. 
Treating these oscillations as rapid relative to all other wavenumber
dependence, we approximate the leading term
$\cos^2{k\so}$ as $1/2$. The large $k\so$ values also leave $\sin^2{k\so}/(k\so)^2$
small relative to $\cos^2{k\so}$ and so we drop this term.
These approximations then yield
\begin{eqnarray}\label{eq:F1D}
F_{\ln \so}&\sim&\int_{0} ^ {\infty} \frac{8\pi^2\Vsur A_0^2}{(P(k)/\Pt+{(n\Pt)}^{-1})^2}
\frac{1}{2}\exp{(-2(k\Sigs)^{1.4})}\exp{(-k^2 \Sign^2)}\frac{4\pi k^2 dk}{2(2 \pi)^3}.
\end{eqnarray}
The resulting error on the location of the baryonic peak is
\begin{eqnarray}
\sig_{\ln \so}&=&\sig_{\so}/\so=\sqrt{F^{-1}_{\ln \so}}\\
&=&
\left[ 
\Vsur A_0^2
\int_{0}^{\infty} dk
\frac{k^2 \exp{(-2(k\Sigs)^{1.4})}\exp(-k^2 \Sign^2)}{\left({P(k)\over\Pt}+{1\over n\Pt}\right)^2}
\right] ^{-1/2}\label{eq:sig1D}.
\end{eqnarray}

The fractional error on the location of the peak from the observed galaxy redshift surveys, $\sig_{\ln \so}$ ($=\sig_{\so}/\so$), is equivalent to the fractional error on the distance estimation when the physical location of the peak, that is, the true value of the sound horizon $s$ at the drag epoch, is well known from the CMB. More
specifically, it is equivalent to the distance information exclusively
from baryonic peaks as a standard ruler, as we have used only $\partial
\PB/\partial \ln \so$ in deriving equation (\ref{eq:F1D}). In comparison
to the full Fisher matrix (eq.\ [\ref{eq:fullFij}]), knowing that $\partial
P(k)/\partial \ln k$ in equation (\ref{eq:fullFij}) is dominated by $\partial
\PB(k)/\partial \ln k$, we hypothesize that equation (\ref{eq:F1D})
is a good approximation to equation (\ref{eq:fullFij}) in spherical symmetry.

\section{2-D approximations of the Fisher matrix}\label{sec:2DA}
In this section, we upgrade the spherically symmetric model in the previous section to a 2-D model. In allowing anisotropic behavior of the correlation function, we want to measure the location of the baryonic peak in the correlation function along ($\spa$) and across the line of sight direction ($\spe$) in the reference cosmology, that is, two axes of an oblate ellipsoidal ridge in the correlation function. The Fourier transform of the ellipsoid is $\sin{x}/x$ where $x=\sqrt{\kpe^2\spe^2+\kpa^2\spa^2}$.

Measuring the fractional errors on $\spe$ and $\spa$ is equivalent to measuring the fractional errors on $\DAA/s$ and $s\hzz$, respectively, where $s$ is the true physical value of the sound horizon. When the precision on the sound horizon $s$ from the CMB data is much better than the precision on  $s\hzz$ and $\DAA/s$ from galaxy redshift surveys, again, the errors on $\spe$ and $\spa$ are virtually equivalent to the errors on $\DAA$ and $\hzz$.

When $\Sigpe <  \Sigpa$, the baryonic peak along the line of sight direction is further broadened. In Fourier space, the modes along the line of sight are damped further than other modes at given $k$, introducing an angular dependence within the integrand of $F_{ij}$.

We choose parameters $p_1=\ln{\spe^{-1}}$ and $p_2=\ln{\spa}$. We choose $\ln{\spe^{-1}}$ instead of $\ln{\spe}$ to be consistent with the sign of derivatives for $\DAA$ in equation (\ref{eq:fullFij}). Using $\ln{\spe}$ instead would only change the sign of the off-diagonal term of the final Fisher matrix we will derive. The derivatives with respect to the anisotropic distances are
\begin{eqnarray}
f_1(\mu) \equiv \partial \ln x/\partial\ln \spe^{-1} &=& \mu^2-1\\
f_2(\mu) \equiv \partial \ln x/\partial \ln \spa &=& \mu^2,  
\end{eqnarray}
when we evaluate the derivatives $f_1$ and $f_2$ at the fiducial cosmology, i.e., the true cosmology where $\spe=\spa=s$.
Using these we can write the 2-D Fisher matrix as
\begin{eqnarray}
F_{ij}&=&\int_{-1}^{1}\int_{0}^{\infty} \frac{\Vsur}{(P(k)R(\mu)+\sn)^2}\left[ \sqrt{8 \pi^2} A_0 \Pt  R(\mu)\frac{\partial P_{\rm BAO}(x)}{\partial \ln x} \right ]^2 \left [ \frac{\partial \ln x}{\partial p_i}\frac{\partial \ln x}{\partial p_j} \right ]\frac{2\pi k^2 dk\;d\mu}{2(2 \pi)^3} \nonumber \\
\end{eqnarray}
where $R(\mu)=(1+\beta\mu^2)^2$ is the linear redshift distortions. 
We can make a similar approximation to equation (\ref{eq:dPbao}) for $\partial \PB(x)/\partial \ln x$:
\begin{eqnarray}
\partial \PB(x)/\partial \ln x &\propto& \partial(\sin{x}/{x})/\partial \ln x \\
&=&\cos{x}-\sin{x}/x. 
\end{eqnarray}
We subsequently approximate that $\cos^2{x} \sim 1/2$, as before. 
\begin{eqnarray} 
F_{ij}&=&\int_{0}^{\infty}  \frac{2\pi k^2 dk }{(2 \pi)^3} 
\frac{\Vsur}{2} \exp{\left [-2(k\Sigs)^{1.4}\right ] }\nonumber \\
&&\int_{-1}^{1}\frac{d\mu}{2}
\frac{(\sqrt{8 \pi^2} A_0 \Pt R(\mu))^2}{(P(k)R(\mu)+\sn)^2} f_i(\mu) f_j(\mu) 
\exp{\left [ -k^2(1-\mu^2)\Sigpe^2-k^2\mu^2\Sigpa^2) \right ] } \\[6pt]
&=& \Vsur A_0^2
\int_{0}^{1} d\mu\; f_i(\mu) f_j(\mu)
\int_{0}^{\infty} dk\,{k^2 \exp\left [-2(k\Sigs)^{1.4}\right ] \over
\left({P(k)\over\Pt}+{1\over n\Pt R(\mu)}\right)^2} 
\exp\left[-k^2(1-\mu^2)\Sigpe^2-k^2\mu^2\Sigpa^2\right] \label{eq:F2D}  \nonumber \\
\end{eqnarray}
where $\Sigs$ is the Silk damping scale, $\Sigpe$ is the real-space
nonlinear Lagrangian displacement, and $\Sigpa$ is the redshift-space
nonlinear displacement along $\hat{z}$. Equation (\ref{eq:F2D}) holds even if the redshift distortion effect $R$ deviates from the linear redshift distortions: e.g., $R(k,\mu)$ (see \S~\ref{subsec:RedD}). The quantity $(n\Pt R)^{-1}$ becomes the effective shot noise $\Neff^{-1}$. Any additive power due to nonlinear growth or bias will decrease $\Neff$ below $n\Pt R$.

We propose to use equation (\ref{eq:F2D}) as our fitting function to the full Fisher matrix calculation (eq.\ [\ref{eq:fullFij}]). As usual, the covariance matrix is the inverse of the Fisher matrix.  Unless $\Sigpe$ or $\Sigpa$ is very large,
the $\mu$ dependence of the integrand is mild and therefore easy to
compute.  We find excellent convergence using 
a simple midpoint method with a grid of 20 points in $\mu$
and 50 points in $k$ with $dk=0.01\ihMpc$ (up to $k=0.50\ihMpc$).
One can precompute $P(k)/\Pt$ at the gridpoints in $k$, as these
do not depend on $n\Pt$, $\beta$, $\Sigpe$, or $\Sigpa$. 
The integral over $k$ can be done once for all three matrix elements,
so that computing the full matrix takes negligibly more time than computing
one element. 

By comparing the numerical results from equation (\ref{eq:F2D}) with those from equation (\ref{eq:fullFij}), which is presented in \S~\ref{sec:result}, we derive $A_0$. We calculate $\Sigs(\equiv 1/\ks)$ from the equation given in \citet{EH98}\footnote{$\ks=1.6(\Ob)^{0.52}(\Oh)^{0.73}\left[1+(10.4\Oh)^{-0.95}\right] h^{-1} (\ihMpc)$}. For WMAP1, $A_0=0.4529$ and $\Sigs=1/\ks=7.76\hMpc$. For WMAP3, $A_0=0.5817$ and $\Sigs=1/\ks=8.38\hMpc$. Note that $\Vsur$ should be in $h^{-3} \Mpc^{3}$ if $k$ is in $\ihMpc$ units. The derived $A_0$ values are consistent with the analytic estimates from equation (\ref{eq:fullP}) although not exact.

Approximating the Silk damping effect as a Gaussian function, that is, $\exp{(-k^2\Sigs'^2)}$, is less exact in the limit of large $\Sign$ and a small $\nPt$ but is convenient because we can add the effect of the Silk damping and the nonlinear damping quadratically (i.e., $\Sig^2=\Sigs'^2+\Sign^2$ in eq.\ [\ref{eq:gausilk}]).  We quote $A'_0$ and $\Sigs'$ for this case as well. For WMAP1, we calculate $A'_0=0.3051$ with a choice of $\Sigs'=8.3\hMpc$, and for WMAP3, $A'_0=0.3794$ with a choice of $\Sigs'=8.86\hMpc$.

\subsection{Redshift distortions and photometric redshift surveys}\label{subsec:RedD}
The contribution of $n\Pt R$ in equation (\ref{eq:F2D}) implies that the effect of redshift distortion $R$ offsets the behavior of the shot noise $\sn$. On large scales, the anisotropic contribution from $R(\mu)$ relative to $R=1$ will increase $\Neff=n\Pt R$ along the line of sight, decreasing the effective shot noise not only for the measurement of $\hzz$ but also for $\DAA$, albeit by a smaller amount. 

If the distance to galaxies are uncertain in an uncorrelated way, e.g., due to thermal peculiar velocities (i.e., the finger-of-God effect) or due to photometric redshifts, it is straightforward to include this effect in equation (\ref{eq:F2D}) by fixing $R(\mu)$: we simply need to include a Gaussian uncertainty that corresponds to halo velocity dispersion due to the finger-of-God effect or redshift uncertainty for photometric surveys. That is,
\begin{eqnarray}\label{eq:Rsig}
R(k,\mu) = (1+\beta\mu^2)^2\exp{(-k^2\mu^2\Sigz^2)}.
\end{eqnarray}
Note that we do not increase values of $\Sigpa$ by the amount of uncertainty in the distance. The reason why we only need to modify $R$, whether due to the nonlinear redshift distortions or due to photometric redshift errors, is because the resulting exponential suppression in power spectrum ($P \rightarrow  P \exp{(-k^2\mu^2\Sigz^2)}$) not only decreases the signal but also decreases the variance from the CDM power spectrum in the line-of-sight direction. The net effect is thus equivalent to a relative increase of shot noise. In \S~\ref{subsec:photoz}, we present distance error estimates from a photometric redshift survey using equation (\ref{eq:Rsig}).

 In the case of photometric redshift errors, there is an additional drawback other than the exponential suppression of power in the line-of-sight direction: features in the transverse power spectrum are smeared by projections of clustering at different redshifts onto a mean redshift.
We ignore this effect in our formulation. As we presented in \citet{SE03}, this effect will increase the error on $\DAA$ by $13\%$ for photometric uncertainty ($1\sig$) of $4\%$ in $1+z$ at $z=1$.

Meanwhile, using $R(\mu)\exp{(-k^2\mu^2\Sigz^2)}$ may be a redundant correction in the case of thermal peculiar velocities, as we may already be including these peculiar velocity term in the computation of $\Sigpa$ (of course, in redshift space) from simulations. In fact, it is more conservative to put this term into $\Sigpa$ than it is to put into $\Sigz$, because while both cases decrease the signal by the same amount, the latter case also decreases the noise by the same amount while the former does not. Therefore, we do not take any steps to isolate thermal velocities in the Lagrangian displacements and move that contribution to $\Sigz$. We simply reserve $\Sigs$ for the inclusion of observational uncertainties such as spectroscopic or photometric redshift errors.

\subsection{A return to spherical symmetry}\label{subsec:symm}
When $\Sigpe =  \Sigpa$ and $R=1$,  the nonlinear exponential damping becomes isotropic, and the 2-D Fisher matrix (eq.\ [\ref{eq:F2D}]) reduces to 1-D Fisher matrix (eq.\ [\ref{eq:F1D}]) multiplied by simple angular integrals.
\begin{eqnarray}
F_{11}&=&F_{\ln \so}\int_{0}^{1}(\mu^2-1)^2 d\mu\\
F_{22}&=&F_{\ln \so}\int_{0}^{1}(\mu^2)^2 d\mu\\
F_{12}&=&F_{\ln \so}\int_{0}^{1}\mu^2(\mu^2-1) d\mu
\end{eqnarray}
This gives a $2\times 2$ matrix of
\begin{equation}\label{eq:F1Dp22}
F_{ij}= F_{\ln \so} \left  (  \begin{array}{cc}
\frac{8}{15} & -\frac{2}{15} \\
-\frac{2}{15} & \frac{3}{15}  \end{array} \right).
\end{equation}
Then the covariance matrix is 
\begin{equation}\label{eq:C1D}
C_{ij}=F^{-1}_{ij}= F^{-1}_{\ln \so} \left  (  \begin{array}{cc}
\frac{9}{4} & \frac{3}{2} \\
\frac{3}{2} & 6  \end{array} \right).
\end{equation}
The off-diagonal term is not zero, implying that the constraints on
$\spe^{-1}$ and $\spa$, and hence the constraints on $\DAA$ and $\hzz$, are not independent
even in the limit of superb data. The off-diagonal term being positive means that $\DAA$ and $\hzz^{-1}$ are anticorrelated. For all modes except those at $\mu=0$ and $\mu=1$, there exist positively correlated changes in $\DAA$ and $\hzz$ that leave the measured quantity of the mode unchanged. Summing all the modes, with relative weight depending on $\mu$, leaves a net anticorrelation between $\DAA$ and $\hzz^{-1}$. Geometrically, we are attempting to measure the axes of an ellipsoidal shell, but a quadrupole distortion of the shell is less well constrained because it leaves the intermediate angles relatively unchanged.  The value of $\roff=C_{12}/\sqrt{C_{11}C_{22}} \approx  0.41$ from equation (\ref{eq:C1D}). In fact, we find that $\roff \approx 0.4$ regardless of the choice of $\Sigpe$, $\Sigpa$, $n\Pt$, and $\beta$.

We recommend that this covariance be included in assessing the 
implications of acoustic scale measurements for dark energy.  We note
that $\DAA$ is an integral of $\hzz^{-1}$, so the fact that the acoustic
scale measurements of these two are anticorrelated means that the 
constraints are slightly stronger than the diagonal errors would imply.

If one is fitting a model in which the transverse and radial distance scales are required to be the same, for example at low redshift, then this implies a contraction of the Fisher matrix with the vector $(1,-1)$. This yields an error $\sig_{\ln \al}=\sig_{\ln \DAA} \sqrt{(1-\roff^2)/(1+2\roff \sig_{\ln \DAA}/\sig_{\ln \hzz}+\sig^2_{\ln \DAA}/\sig^2_{\ln \hzz})}$ where $\sig^2_{\ln \DAA}=C_{11}$ and $\sig^2_{\ln \hzz}=C_{22}$. For the case of spherical symmetry, this reduces to $(2/3)\sig_{\ln \DAA}$ $(={F^{-1/2}_{\ln \so}})$ as expected from equation (\ref{eq:C1D}).

\section{Testing the approximation}\label{sec:result}
\subsection{Distance errors from the full Fisher matrix and the 2-D model}
We next compare the forecasts on $\DAA$ and $\hzz$ from the full Fisher
matrix (eq.\ [\ref{eq:fullFij}], hereafter, `full-D errors') to
those from our fitting formula (eq.\ [\ref{eq:F2D}], hereafter, `2-D errors') for various values of
$n\Pt$, $\Sigpe$, and $\Sigpa$.  We hold $R=1$, as this is a minor
term. We will present the comparisons for the WMAP3 cosmology; the performance is similarly good for the WMAP1 cosmology.

We first study the case where $\Sigpa=\Sigpe$. The upper panels of Figure \ref{fig:2Dc} show expected errors on $\DAA$ and $\hzz$ using the baryonic peaks as a function of nonlinear parameter, $\Sigpe$ and $\Sigpa$ respectively, for various $\nPt$. The blue lines show the full-D errors and the black lines show the 2-D errors. The figure shows that the errors from our fitting formula are in excellent agreement with the errors from the full Fisher matrix calculations. Deviations are typically only a few \%. The deviation becomes only noticeable for very large shot noise $n\Pt$ ($< 0.1$), and very large $\Sigpa$ ($> 20 \hMpc$) where our assumptions to derive equation (\ref{eq:F2D}) break down.

We next present the distance errors when $\Sigpe < \Sigpa$. As nonlinear redshift distortions elongate the Lagrangian displacement distribution of pairs at $100\hMpc$ along $\hat{z}$ by $(1+f)$, where $f=\Omega^{0.6}$, the distance errors for $\Sigpe < \Sigpa$ represent the effect of nonlinear redshift distortions on baryonic features. We consider cases where $c=\Sigpa/\Sigpe=2$ and $3$, while $c=2$ corresponds to more realistic redshift distortions, and $c=3$ depicts an extreme case of anisotropy in Lagrangian displacement fields. 

In the lower panels in Figure \ref{fig:2Dc}, we show the full-D errors and the 2-D errors from equation (\ref{eq:F2D}) for $c=2$.
Again. the 2-D model reproduces the full-D errors to a good extent. We find a similarly good agreement between the full-D errors and the 2-D errors for $c=3$. 

Note that errors from the fitting formula predict $\roff \sim 0.4$ regardless of $c$ and $R(\mu)$. Values of $\roff$ from the full-D  calculations are fairly close to the expected values.

\subsection{Photometric redshift surveys}\label{subsec:photoz}
We use our fitting formula to derive distance errors for photometric redshift surveys. We do this by modifying $R(\mu) \rightarrow R(k,\mu)$ as described in \S~\ref{subsec:RedD}. We use WMAP3 cosmology and assume a photometric redshift error ($1\sig$) of $\Sigz=34\hMpc$, which corresponds to $1\%$ rms in $(1+z)$ at $z=1$. We use equation (\ref{eq:Rsig}) to properly include the photometric redshift errors into the Fisher matrix. Figure \ref{fig:photoz} shows good agreement between errors on $\DAA$ from our fitting formula and errors from the full-D Fisher matrix. Although the discrepancy is larger than spectroscopic cases (i.e., Figure \ref{fig:2Dc}), the offset is at most 8\%. Fortunately the deviation happens  to be small at the common values of $\nPt$ for photometric redshift surveys, e.g., $\nPt=3$ to $10$.

\section{Comparing the distance estimates to N-body data}\label{sec:Nbody}
In this section, we compare our revised full Fisher matrix formalism (eq.\ [\ref{eq:fullFij}]) to the distance estimates from a $\chi^2$ analysis of N-body data. We find that our revised Fisher matrix formalism provides an excellent forecast for the distance errors from the N-body data.

We use the N-body simulations from \citet{SE05} and perform a $\chi^2$ analysis  to extract the acoustic scale from the simulated power spectrum. \citet{SE05} used WMAP1 cosmology and the Hydra code \citep{Couchman95} to generate 51 sets of cosmological N-body simulations with a box size of $512^3\trihMpc$ that were evolved from $z=49$ to $z=3$ (30 sets at $z=3$), $z=1$, and $z=0.3$. In the $\chi^2$ analysis, the spherically averaged real-space power spectra $P_{\rm obs}(\kref)$ from the N-body data are fitted to model power spectra including a scale dilation parameter $\al$, $P_{m}(\kref/\alpha)$. In detail, we use an additive polynomial function to represent the effect of shot noise and nonlinearity:
\begin{equation}\label{eq:chi}
P_{\rm obs}(k_{\rm ref})=(b_0+b_1 k_{\rm ref})\times P_{m}(k_{\rm ref} /\alpha)+(a_0+a_1 k_{\rm ref}+a_2 k_{\rm ref}^2).
\end{equation}
The additive polynomial function should also suppress the distance information from the broad-band shape of the power spectrum. Recall that, in the Fisher matrix calculations, we explicitly subtracted from the Fisher matrix the distance information from the broadband shape (\S~\ref{sec:full}).

\citet{SE05} used the linear power spectrum for $P_{m}(\kref/\alpha)$; we improve this by modifying the linear power spectrum to include the nonlinear erasure of the acoustic peaks based on \citet{ELag06}. In detail, we use
\begin{equation}\label{eq:Psm}
P_m=(P_{\rm linear}(k)-P_{\rm smooth}(k))\exp{\left[-k^2\Sign^2/2\right ]}+P_{\rm smooth}(k),
\end{equation}
where $P_{\rm smooth}$ is the ``no wiggle'' form from \citet{EH98}. $\Sign$ here does not need to be precisely the same as $\Sign$ measured in \citet{ELag06} and used in the Fisher matrix calculations. $\Sign$ is used here merely to improve the template power spectrum. The resulting errors on $\al$ presented below are a smooth function of our choice of $\Sign$ for $P_m$.

 In equation (\ref{eq:chi}), the fit parameters are $\al$, a multiplicative bias $b_0$, a scale-dependent bias $b_1$, and additive terms $a_0$, $a_1$, and $a_2$ from nonlinear growth, bias, or shot noise. The dilation parameter $\al$ is the ratio of the true distance to our estimate from a $\chi^2$ analysis. For simplicity, we set the true cosmology as the reference cosmology, and therefore the mean value of $\al$ is expected to be unity. The error on $\alpha$ represents the combined errors on $\DA$ and $\hz$, more specifically $\sigma_\al=(2/3)\sig_{\DAA}$ (\S~\ref{subsec:symm}), as we use the power spectra in real space.

Since we do not know the true covariance matrix of power spectrum, we do not trust differences in $\chi^2$ to give an accurate error on $\al$. Instead, we calculate the mean value and error of $\alpha$ using jackknife subsampling of simulations while assuming an error on the band power to be given by a Gaussian random field assumption, i.e., based on the number of modes contributing to the band power. We derive 51 (30) subsamples by removing one simulation each time from our 51 (30 at $z=3$) simulations. Each subsample is then fitted assuming a Gaussian random error on the band power. The error and mean value of $\alpha$ is computed from the variations among the jackknife subsamples. As we emphasized in \citet{SE05}, assuming a Gaussian error ignores mode-coupled errors between wavenumbers and therefore underestimates the statistical noise on small scales. Nevertheless, the variations among jackknife sampling should reflect the true non-Gaussian error to a reasonable extent, as these subsamples are drawn from actual nonlinear N-body data. That is, $\chi^2$ statistic slightly misweights the data on small scales in each subsample but the variation among subsamples should not produce an overly optimistic $\sigma_\al$ compared to the true error.

The resulting $\sigma_\al$ from the revised $\chi^2$ analysis is, in general, similar to or slightly smaller than the quoted values in \citet{SE05}. At all redshifts, the value of $\sig_\al$ is stable with respect to different fitting ranges of wavenumber $k<\kfit$ ($\kfit=0.3$, $0.4$, and $0.5\ihMpc$) or the inclusion of $b_1$. Note that in \citet{SE05}, we had to use a narrower fitting range ($\kfit=0.3\ihMpc$) at $z=0.3$. We will quote errors for $\kfit=0.4\ihMpc$ without $b_1$. When fitting to the matter power spectrum, we will use $\nPt \sim 80$.

Meanwhile, in all cases but one, the mean values of $\al$ are within $1.44 \sig_\al$ of unity, including 17 out of 24 cases within $1\sig_\al$, indicating that there is no detection of a bias on the mean value of $\al$. The worst out of 24 cases, without $b_1$ with $\kfit=0.5\ihMpc$ at $z=0.3$, gives a mean value of $\al$ that is $2.24\sig_\al$ off from unity. In detail, in \citet{SE05}, we reasoned that using $b_1$ will slightly bias $\al$ above unity, beyond $1\%$ in worse cases, because the fitting process will favor a negative $b_1$ to match the erased portion of the BAO, and the resulting phase shift in the baryonic peaks will be compensated by $\al$ above 1. Now that we have accounted for the erased features in generating $P_m(k)$, this bias on $\al$ has decreased compared to \citet{SE05}.

At $z=3$, we find $\sigma_\al \sim 0.35\%$ for $\Vsur=4\trihGpc$ from the N-body data. This corresponds to $0.7\%$ for $\Vsur=1\trihGpc$. The full-D error ($c=1$, $\Sign=3.07\hMpc$ and $\nPt \sim 80$) predicts $\sigma_\al=0.65\%$, which is about $8\%$ smaller than the N-body data. Note that these Fisher matrix errors are different from the values quoted in \citet{SE05}, mainly because of the different fiducial cosmology.

At $z=1$, we find $\sigma_\al = 0.38\%$ for $\Vsur=6.845\trihGpc$. For $\Vsur=1\trihGpc$, this rescales to $\sigma_\al = 0.99\%$. The corresponding full-D error for $\Sign=5.90\hMpc$ and $\nPt \sim 80$ predicts $\sigma_\al = 1.01\%$ for $\Sign=5.90$ and $\nPt \sim 80$, in excellent agreement.

At $z=0.3$, we calculate $\sigma_\al =0.60\%$ for $\Vsur=6.845\trihGpc$ of simulations. For $\Vsur=1\trihGpc$, the error rescales to $\sigma_\al =1.57\%$. In comparison, the full-D error for $\Sign=8.15\hMpc$ and $\nPt \sim 80$ predicts $\sigma_\al = 1.50\%$.

We also compute the results when the baryonic signature in these nonlinear density fields at $z=0.3$ is reconstructed using the simple scheme presented in \citet{ESSS06}. We use a $10\hMpc$ Gaussian filter to smooth gravity, displace the real particles and a set of smoothly distributed reference particles by the linear theory motion predicted from the nonlinear density fields, find new density fields from the difference of the density fields of the real particles and the reference particles, and compute the power spectra of these new fields. Fitting these spectra, we find $\sig_\al$ of $0.34\%$ for $4\trihGpc$ of simulation volume\footnote{We only include 30 out of 51 sets of simulations because the other 21 sets were produced from initial input power spectra that omitted the baryonic signature at $k> 0.3\ihMpc$. Hence, we use the full 51 only when the non-linearities have reduced the role of the $k > 0.3\ihMpc$.} and thus $0.68\%$ for $1\trihGpc$. This is a considerable improvement over the 1.57\% measured without reconstruction. This matches the Fisher matrix prediction if $\Sign\sim 3.4\hMpc$, while we measured $\Sign \sim 4.4\hMpc$ in \citet{ESSS06}. Whether this difference is due to sample variance or due to our technical difficulties in estimating $\Sign$ for the reconstructed density fields remains to be studied. While this result is an example for a negligible shot noise without galaxy bias, our $\chi^2$ result implies an impressive prospect for the reconstruction in the existing and future galaxy surveys. We will investigate the effects of shot noise and galaxy bias on reconstruction in future papers.

For the biased case at $z=0.3$, we use the MASS case with $m=10$ from \citet{SE05}, which has galaxy bias similar to the LRG samples in SDSS \citep{Zehavi05a}. From the $\chi^2$ analysis, we find $\sigma_\al =0.74\%$ for $\Vsur=6.845\trihGpc$. The error is equivalent to $\sigma_\al =1.94\%$ for $\Vsur=1\trihGpc$.  
To evaluate errors using the Fisher matrix, we need to know the value of $\nPt$ that corresponds to this case. This is tricky because the nonlinear growth and bias effect in the MASS case with $m=10$ increases the small-scale power above a nominal shot noise that would be strictly from the inverse number density. In addition, this anomalous power is not likely constant in scale. Thus, we roughly estimate $\Neff=\nPt$ using the best fit additive polynomial in equation (\ref{eq:chi}) evaluated at $k=0.2\ihMpc$: we derive $\Neff \sim 1.5$, which corresponds to $\sigma_\al \sim 2.2\%$ for $\Vsur=1\trihGpc$.

In general, we find excellent agreement between the errors from the $\chi^2$ analysis of the N-body data and the analytic full-D/1-D errors. The discrepancy is at most $13\%$ for the cases we have shown. Some of the discrepancy is probably from sample variance. With only 51 values of $\al$, we cannot estimate $\sig_\al$ perfectly. In addition, while the small-scale power added due to various nonlinearities can be accounted for by properly modifying $\Neff$ below $\nPt$, variance due to nonlinearity are not accounted for in our Fisher matrix formalism: this will induce slightly larger distance errors from the N-body data than the full-D/1-D errors. While our fitting formula in the $\chi^2$ analysis produces reasonable estimates of distance errors, it is not necessarily a most optimized one \citep[other examples can be found in ][]{Huff06,Koehler06,Smith06}. More sophisticated fitting schemes may also reduce the discrepancy between the $\chi^2$ analysis and the Fisher matrix estimates.

\section{Discussion}\label{sec:Discussion}

We have computed the cosmological distance errors available from the BAO in future galaxy redshift surveys using a Fisher matrix formalism that incorporates the Lagrangian displacement field to account for the erasure of the BAO due to nonlinear growth, bias, or redshift distortions. The resulting formalism is only a function of survey volume, shot noise, and a nonlinear parameter that can be measured quantitatively. We have derived physically motivated lower dimensional approximations to the full Fisher matrix and showed excellent agreement between distance error estimates from the approximations and the full Fisher matrix. We present the resulting fitting formula to calculate a 2-dimensional covariance matrix for $\DA$ and $\hz$. The fitting formula straightforwardly applies to photometric redshift surveys with a simple modification to the effective shot noise, although its agreement to the estimates from the full Fisher matrix becomes slightly degraded. The merit of the fitting formula is its simplicity in terms of input variables and computation. 

Finally, we compared the error estimates from the revised Fisher matrix with the error estimates from a $\chi^2$ analysis of N-body simulations. For the $\chi^2$ analysis, we also used the Lagrangian displacement field to account for the nonlinear effect on baryonic peaks in the template power spectrum. This improved various aspects of the error estimates. We showed that both error estimates are in excellent agreement. The discrepancies (at most $13\%$) could be due to non-Gaussianity contributions to the variance not included in the Fisher matrix calculation, but may simply be sample variance due to the limited number of N-body simulations.

At \url{http://cmb.as.arizona.edu/~eisenste/acousticpeak/bao\_forecast.html}, we provide a C-program for the fitting formula that can be used both for spectroscopic and photometric redshift surveys.

To use the formula, one must construct $\Sigpe$, $\Sigpa$, and $nP$ at the redshift of the survey in question.  $\Sigpe$ and $\Sigpa$ are
dominated by the bulk flows in the Universe; their simple scalings were
given in \S~\ref{sec:full}, although one should probably impose a lower
$\sigma_{\rm 8,matter}$ and rescale $\Sig_0$ for WMAP3 cosmology. \citet{ELag06} found that highly biased tracers could 
increase the Lagrangian variances by small amounts, although this has not been calibrated in detail. The value of $\nPt$ is easy to calculate by using $\Pt=2710\sig^2_{8,g}$ for WMAP3, and $\Pt=2875\sig^2_{8,g}$ for WMAP1.

\begin{figure}
\plotone{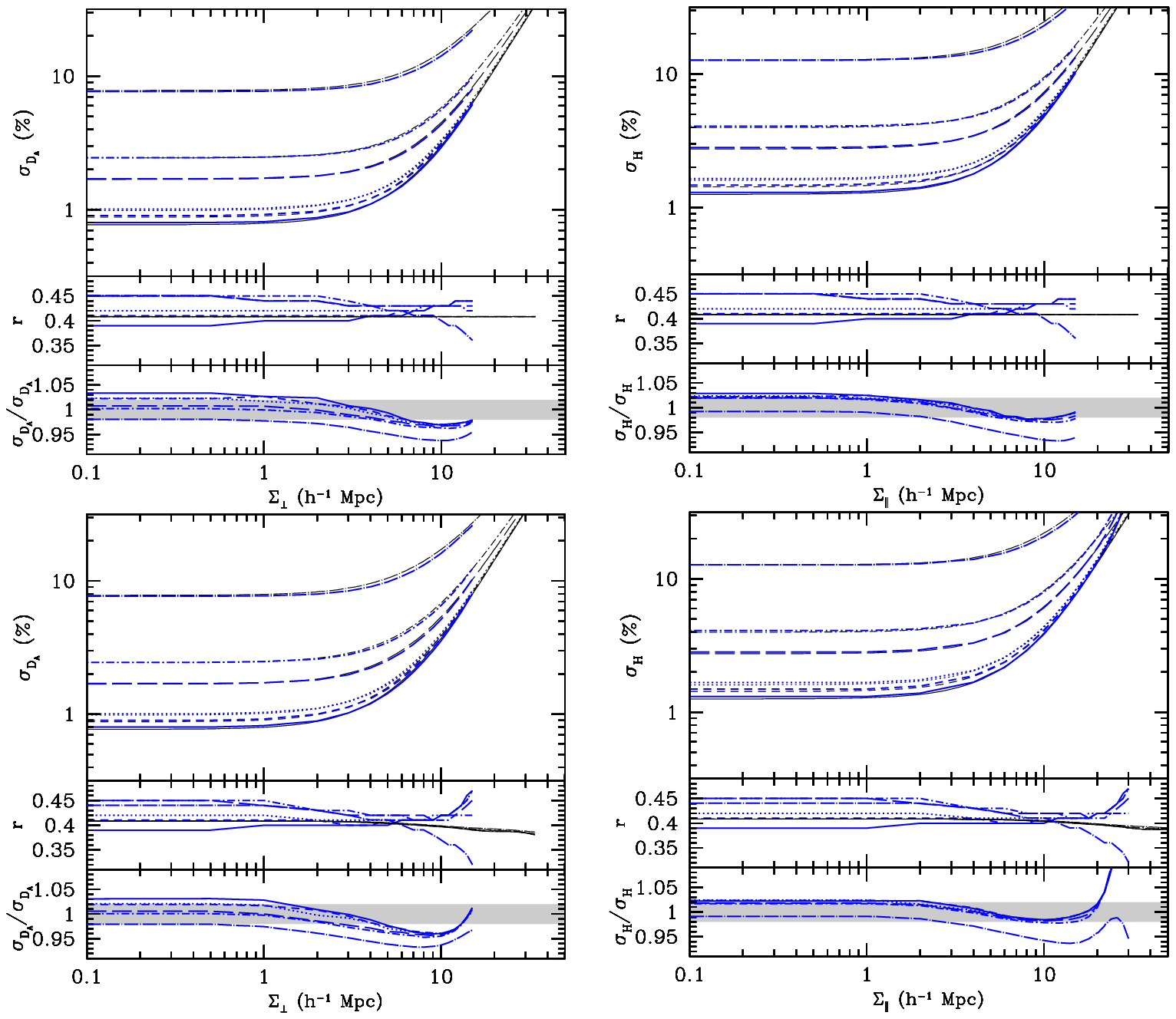}
\caption{2-D errors from equation (\ref{eq:F2D}) (black lines) and errors from the full Fisher matrix calculations (blue lines) for WMAP3. Upper panels: $c=\Sigpa/\Sigpe=1$. Lower panels : $c=2$. Left: distance errors on $\DAA$. Right: distance errors on $\hzz$. Off-diagonal terms in the middle field of each panel are defined as $\roff=C_{12}/\sqrt{C_{11}C_{22}}$. The bottom field of each panel shows the discrepancy between the 2-D errors and the full-D errors as a ratio of the two. The shaded region corresponds to 2\% of discrepancy. We find that the 2-D model gives excellent fits to the errors from the full Fisher matrix calculations. Solid lines : $n\Pt=76.1$, short-dashed : $n\Pt=7.61$, dotted : $n\Pt=3.81$, long-dashed : $n\Pt=0.761$, dot-short-dashed : $n\Pt=0.381$, and dot-long-dashed : $n\Pt=0.076$. These values are chosen because $\nPt\sim 0.76$ is appropriate for the Luminous Red Galaxy sample from SDSS in real space. }
\label{fig:2Dc}
\end{figure}

\begin{figure}
\epsscale{0.5}
\plotone{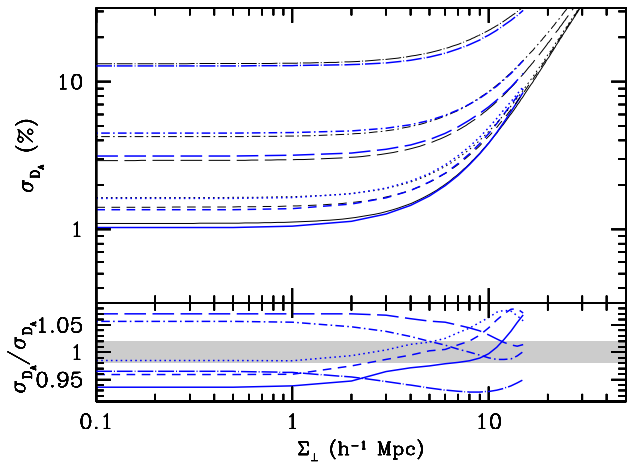}
\caption{Errors on $\DAA$ for photometric redshift surveys. We assume a redshift error of $\Sigz=34\hMpc$. We compare the 2-D errors from equation (\ref{eq:F2D}) using $R(k,\mu)$ in equation (\ref{eq:Rsig}) (black lines) and the errors from the full Fisher matrix calculations (blue lines) for WMAP3. The lower field of the panel shows the discrepancy between the 2-D errors and the full-D errors as a ratio of the two. The 2-D errors from our fitting formula are in good agreement with the full-D errors: the discrepancy is at most $8\%$ but smaller for $n\Pt=3-10$. Solid lines : $n\Pt=76.1$, short-dashed : $n\Pt=7.61$, dotted : $n\Pt=3.81$, long-dashed : $n\Pt=0.761$, dot-short-dashed : $n\Pt=0.381$, and dot-long-dashed : $n\Pt=0.076$. }
\label{fig:photoz}
\epsscale{1}
\end{figure}

Reconstructing the density field reduces  the Lagrangian displacements
and restores the baryonic peaks from nonlinear degradation 
\citep{ESSS06}.  In \citet{ESSS06}, we found that our simple scheme was
able to decrease $\Sigpe$ and $\Sigpa$ about by half at $z=0.3$ (from $\Sigpe \sim 8\hMpc$ and $\Sigpa \sim 14\hMpc$). In the middle panel of Figure
\ref{fig:2Dc}, this decrease due to the reconstruction happens where
the performance is improving rapidly with $\Sig$, making future galaxy
redshift surveys more promising. The $\chi^2$ results in \S~\ref{sec:Nbody} implies that the improvements may be better than 50\%. We have not yet quantified
these improvements in $\Sigma$ as a detailed function of $nP$ and redshift,
but we recommend, to be conservative, that one consider a 50\% drop in $\Sigma$ as an
estimation of what reconstruction can do; Figure 1 of \citet{ESSS06} estimates that correcting for bulk flow on scales larger than $k\approx0.1\ihMpc$ will decrease the displacement by 50\%. 

Although equation (\ref{eq:F2D}) appears complicated in that one must compute
three two-dimensional integrals, it is far simpler than the full Fisher
matrix. The 12-dimensional problem requires that one compute many integrals
of oscillatory integrands to high precision to avoid degeneracies. In
contrast, the few integrals in the approximation are all smooth and can
be computed rapidly and robustly. Meanwhile, the approximation includes
the anisotropic effects of redshift distortions in the shot noise and
the nonlinear degradations of the acoustic scale.

The 1-D model presented in \S~\ref{sec:1DA} offers an estimate of how the 
performance scales with the non-linear degradation.  Let us rework
equation (\ref{eq:sig1D}), replacing the Silk damping form with a Gaussian 
and approximating the denominator as a power-law $k^{2n}$.  The
performance then becomes proportional to 
\begin{equation}
\int_0^\infty dk\;k^{2-2n} \exp(-k^2 \Sigma^2)
\end{equation}
where $\Sig^2 = \Sigs^2 + \Sign^2$.
Hence, we find that the distance-scale performance scales as 
$\Sig^{1.5-n}$.  
For the case of white noise, we have $n=0$ and the result that 
the precision scales as $\Sig^{3/2}$.  This is familiar with white noise: 
a peak has a higher signal-to-noise ratio proportional to the 
inverse square root of its width, while the measurement of its
centroid scales as the width divided by the signal-to-noise ratio.

However, the cold dark matter power spectrum is better approximated
by $k^{-1}$ near $k\approx 0.15\ihMpc$.  Therefore, if our acoustic
scale measurements are sample-variance limited, we expect to use
$n=-1$, which implies a precision scaling as $\Sig^{5/2}$.
Alternatively stated, the survey volume to reach a given precision
scales as $\Sig^5$.  In the
language of centroiding a peak, what is happening is that the noise
is not white and instead has large correlations between neighboring
separations.  Shifts in the centroid of a narrower peak require larger
changes on smaller scales that the noise model disfavors.

This rapid scaling with $\Sig$ implies that one gains rapidly with
improvements in reconstruction so long as $\Sign$ is not much
smaller than the Silk damping scale of $8\hMpc$.  The difference between
$\Sign=8\hMpc$ and $4\hMpc$ is a factor of 1.8 in distance and
3.2 in survey area.  It should be noted that the redshift-space boosts
of the displacements along the line of sight cause $\Sigpa$
to be interestingly large even at $z\sim 2$.
 
\begin{figure}
\plotone{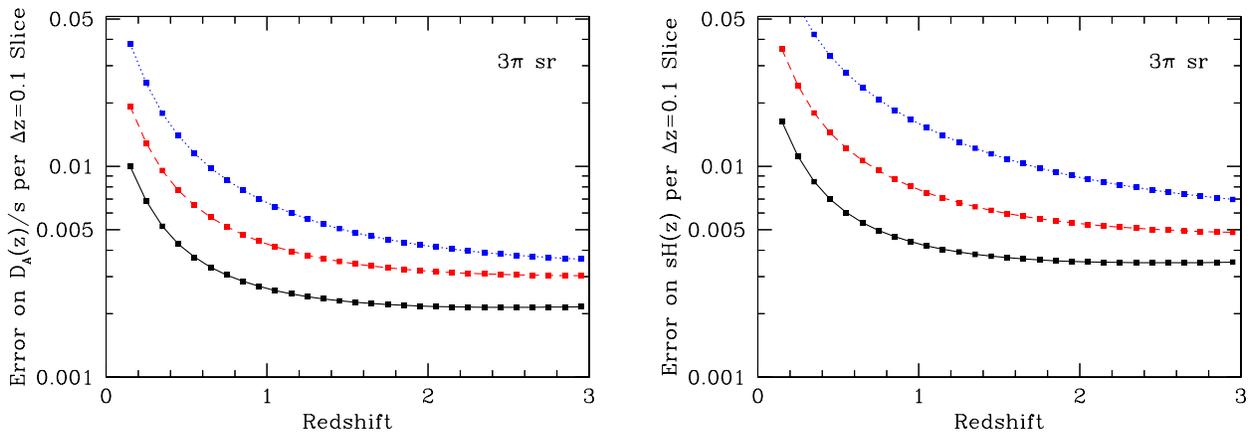}
\caption{The fractional errors on $\DAA/s$ ({\it left}) and $sH$ ({\it right}) available as a function of redshift assuming redshift bins $\Delta z=0.1$ 
and a $3\pi$~sr survey area.  The bottom line in each case is the
cosmic variance limit, assuming perfect linearity and no shot noise.
The top line assumes unreconstructed level of non-linearity and a
shot noise level of $\nPt=3$.  The middle line uses the same
shot noise and assumes that reconstruction can halve the values 
of $\Sigma_\perp$ and $\Sigma_\parallel$.}\label{fig:2Dcv}
\end{figure}

Finally, we consider the cosmic variance limits of the acoustic
oscillation method.  Figure \ref{fig:2Dcv} presents the fractional errors on $\DAA/s$ and $s\hzz$ from a survey of $3\pi$ steradians, i.e., the reasonably accessible
extragalactic sky.  Redshift bins of width 0.1 are used.  The bottom lines
show the precision available for a survey with perfect reconstruction and
no shot noise.  The top lines show the precision for a survey with the 
unreconstructed non-linear degradations and shot noise of $\nPt=3$.
The middle lines show a survey with reconstruction halving the values
of $\Sigma_\perp$ and $\Sigma_\parallel$, again with $\nPt=3$.
This last case is what we would suggest as a reasonable estimate for
a densely sampled survey.  One sees that the acoustic oscillation
distance scale can reach precisions of about 0.4\% and 0.7\% on $\DAA/s$
and $s\hzz$, respectively, for each $\Delta z=0.1$ bin at $z\approx1$,
improving slightly toward $z=3$.  Of course, most dark energy models 
predict smooth trends in $\DAA$ and $\hzz$ on scales of $\Delta z=0.1$, so
in practice one would combine several bins in constructing tests.
Hence, effective precisions better than 0.2\% in distance are available
with the acoustic oscillation method.

\acknowledgements
We thank Martin White and Nikhil Padmanabhan for useful discussions. 
HS and DJE were supported by grant AST-0407200 from the National Science Foundation. HS was supported in part by the National Optical Astronomy Observatory
and Gemini Observatory through Gemini Work Scope No. 0525280-GEM03021.
The Gemini Observatory is operated by the Association of Universities
for Research in Astronomy, Inc., under a cooperative agreement with the
NSF on behalf of the Gemini partnership: the National Science Foundation
(United States), the Particle Physics and Astronomy Research Council
(United Kingdom), the National Research Council (Canada), CONICYT
(Chile), the Australian Research Council (Australia), CNPq (Brazil)
and CONICET (Argentina). DJE was further supported by a Alfred P.\
Sloan Research Fellowship.

\end{document}